\documentstyle[preprint,eqsecnum,prd,aps]{revtex}
\begin{document}
\preprint{
\vbox{\halign{&##\hfil\cr
		& OHSTPY-HEP-T-96-010 \cr
		& NUHEP-TH-96-2       \cr
                & hep-ph/9604237      \cr
		& April 1996          \cr}}}
		
\title{Helicity Decomposition for Inclusive $J/\psi$ Production}

\author{Eric Braaten}
\address{Physics Department, Ohio State University, Columbus OH 43210}

\author{Yu-Qi Chen}
\address{Department of Physics and Astronomy, Northwestern University,
	Evanston IL 60208}

\maketitle
\begin{abstract}

We present a general method for calculating inclusive cross sections 
for the production of heavy quarkonium states with definite polarization
within the NRQCD factorization approach.  Cross sections for polarized 
production can involve additional matrix elements that do not contribute to 
cross sections for unpolarized production.  They can also 
include interference terms between parton processes that produce
$Q \bar Q$ pairs with different total angular momentum.
The interference terms cancel upon summing over polarizations.
Our method can be generalized to $N$ dimensions and is therefore
compatible with the use of dimensional regularization to calculate
radiative corrections.
We illustrate the method by applying it to the production of $J/\psi$
via the parton processes $q \bar q \to c \bar c$ and $g g \to c \bar c$.

\end{abstract}

\vfill \eject

\narrowtext

\section{Introduction}

Calculations of inclusive production rates
of heavy quarkonium have recently been placed on a firm theoretical foundation
by the development of a factorization approach based on 
nonrelativistic quantum chromodynamics (NRQCD) \cite{BBL}.  
In this formalism, the cross section is 
expressed as a sum of products 
of short-distance coefficients and NRQCD matrix elements.
The short-distance coefficients
can be calculated as perturbation series in the coupling constant 
$\alpha_s$ at the scale of the heavy quark mass. The matrix elements
scale in a definite way with the typical relative velocity $v$ of the 
heavy quark in the quarkonium state.  Thus the
production cross sections 
can be calculated systematically to any desired order in $\alpha_s$ 
and $v^2$ in terms of well-defined NRQCD matrix elements.
There have been many recent applications of this formalism to quarkonium 
production in various high energy processes \cite{BFY}.

Most calculations of quarkonium production have been carried out 
using a covariant projection formalism developed for calculations
in the color-singlet model \cite{KKS}.  In the amplitude for producing 
a $c \bar c$ pair with total momentum $P$ and relative momentum $q$, 
the spinor factor $u \bar v$ is replaced by an appropriate Dirac matrix
that projects out a state with the desired angular-momentum quantum numbers.
The resulting prescription is relatively simple for S-wave states.
The Dirac matrix is proportional to $\gamma_5 (\rlap/{\! P} - 2 m_c)$ for a 
$^1S_0$ state and $\not \! \epsilon (\rlap/{\! P} - 2 m_c)$ for a $^3S_1$ state,
and the relative momentum $q$ is set to zero.  For P-waves, the 
Dirac matrix is more complicated, and the amplitude must be differentiated 
with respect to the relative momentum.  The resulting expression for the 
amplitude can be very complicated.

The covariant projection method also has a number of other drawbacks.  
For each angular momentum state, a separate calculation of the cross section 
is required, beginning at the level of the amplitude for $c \bar c$
production.  Another drawback is that
it is difficult to generalize the projection method to 
allow the calculation of relativistic corrections to the cross section.
There is also a potential difficulty in calculating radiative corrections 
using this method.  The Dirac matrices that are used to project onto the 
appropriate angular momentum states are specific to 3 space dimensions.
However the most convenient method for regularizing the ultraviolet and 
infrared divergences that arise in higher order calculations is dimensional 
regularization.  It is not easy to generalize the projection matrices to $N$ 
dimensions, since the representations of the rotational symmetry group are 
different for each integer value of $N$.

In this paper, we develop a new method for calculating quarkonium production 
rates that fully exploits the  NRQCD factorization framework.
The short-distance coefficients are calculated in a form that 
holds for every quarkonium state.  The corresponding NRQCD matrix elements 
are then simplified using 
rotational symmetry and the approximate heavy-quark spin symmetry of NRQCD.
It is only at this stage that the angular momentum quantum numbers of the 
quarkonium state come into play.  Relativistic corrections are easily 
calculated in this framework.   
Our method for calculating the short-distance coefficients
is readily generalized to $N$ spacial dimensions, so that 
dimensional regularization can be used to cut off infrared and 
ultraviolet divergences.  After removing the divergences,
one can specialize to $N=3$ and use rotational 
symmetry and spin symmetry 
to simplify the NRQCD matrix elements.
This approach allows the consistent use of
dimensional regularization to calculate 
inclusive heavy quarkonium production and decays.

Our method has important implications for the production of quarkonium states
with definite polarization.  We demonstrate that cross sections for polarized 
production involve new matrix 
elements that do not appear in cross sections for unpolarized production.
Thus, measurements of cross sections for unpolarized production are not 
necessarily sufficient to predict production rates for polarized 
quarkonium states.
We also show that cross sections for polarized production
can involve interference between parton processes that produce
$c \bar c$ pairs with different total angular momentum $J$.
The interference terms vanish upon summing over polarizations.
Thus the cross sections for producing $c \bar c$ states with definite 
total angular momentum are not sufficient to determine
all the short-distance coefficients in cross sections for polarized production.

The NRQCD factorization approach is summarized in section II.
In section III, 
we present a general matching procedure 
for calculating the short-distance coefficients 
in the factorization formula.  We illustrate the 
method by applying it to 
the parton processes $q \bar q \to c \bar c$ and $g g \to c \bar c$.
The resulting expressions for the cross sections hold for every quarkonium 
state.  In section IV, we show how the NRQCD matrix elements can be 
simplified by using rotational symmetry and the approximate heavy-quark spin 
symmetry of NRQCD.  For simplicity, we focus on the matrix elements for 
$J/\psi$ production.  In section V, we calculate the cross section  
and the spin alignment of the $\psi$ from the parton processes
$q \bar q \to c \bar c$ and $g g \to c \bar c$.  We point 
out that several previous calculations of quarkonium production need to be 
reconsidered in the light of our results.

\section{NRQCD Factorization Formalism}

We consider the inclusive production of a quarkonium state $H$ 
with momentum $P$ and helicity $\lambda$ via a parton process
of the form $1 2 \to H(P,\lambda) + X$.  The 
differential cross section, summed over additional final states $X$, 
can be written
\begin{eqnarray}
\sum_X d \sigma(1 2 \to H(P,\lambda) + X) & &
\nonumber \\
& & \hspace{-1in}
\;=\; {1 \over 4 E_1 E_2 v_{12}} \; {d^3P \over (2 \pi)^3 2 E_P} \;
\sum_X  (2 \pi)^4 \delta^4(k_1 + k_2 - P - k_X) \;
	| {\cal T}_{1 2 \to H(P,\lambda)+X}|^2 ,
\label{sig-M}
\end{eqnarray}
where $E_P = \sqrt{M_H^2 + {\bf P}^2}$ and the sum over $X$ includes 
integration over the 
Lorentz-invariant phase space for the additional particles.

The NRQCD factorization formalism can be used to factor the cross section 
(\ref{sig-M}) into short-distance coefficients and long-distance 
matrix elements \cite{BBL}:
\begin{equation}
\sum_X d \sigma(1 2 \to H(P,\lambda) + X) 
\;=\; {1 \over 4 E_1 E_2 v_{12}} \; {d^3P \over (2 \pi)^3 2 E_P} \;
	\sum_{mn} C_{mn}(P,k_1,k_2) 
	\langle {\cal O}_{mn}^{H(\lambda)} \rangle .
\label{sig-NRQCD}
\end{equation}
The coefficients $C_{mn}$ are functions of the kinematic variables
$P$, $k_1$, and $k_2$.  They take into account the effects of short 
distances of order $1/m_c$ or smaller, 
and therefore can be calculated as perturbation series in the QCD coupling 
constant $\alpha_s(m_c)$.  The matrix elements 
$\langle {\cal O}_{mn}^{H(\lambda)} \rangle$ are expectation values in the 
NRQCD vacuum of local 4-fermion operators that have the structure
\begin{equation}
{\cal O}_{mn}^{H(\lambda)}
\;=\; \chi^\dagger {{\cal K}'}_m^\dagger \psi \; 
	{\cal P}_{H(\lambda)} \;
	\psi^\dagger {\cal K}_n \chi ,	
\label{O-NRQCD}
\end{equation}
where $\psi$ and $\chi$ are the field operators for the heavy quark and 
antiquark in NRQCD, and ${\cal K}_n$ and ${{\cal K}'}_m^\dagger$
are products of a color matrix ($1$ or $T^a$), 
a spin matrix ($1$ or $\sigma^i$), and a polynomial in the gauge covariant 
derivative ${\bf D}$.  The projection operator
${\cal P}_{H(\lambda)}$ can be written
\begin{equation}
{\cal P}_{H(\lambda)} 
\;=\; \sum_S | H({\bf P} = 0,\lambda) + S \rangle 
	\langle H({\bf P} = 0,\lambda) + S | .	
\label{P_H}
\end{equation}
The sum is over soft hadron states $S$ whose total energy is less than the 
ultraviolet cutoff $\Lambda$ of NRQCD.  Thus this operator 
projects onto the subspace of states which in the asymptotic future 
include the quarkonium state $H(\lambda)$ at rest plus soft hadrons.
The normalization of the 
meson states in (\ref{P_H}) must coincide with those in the 
T-matrix element in (\ref{sig-M}).  The standard relativistic 
normalization is
\begin{equation}
\langle H({\bf P}',\lambda') | H({\bf P},\lambda) \rangle 
\;=\; 2 E_P \; (2 \pi)^3  \delta^3({\bf P} - {\bf P}') 
	\; \delta_{\lambda \lambda'}.	
\label{norm_H}
\end{equation}
With this normalization of states, the projection operator 
${\cal P}_{H(\lambda)}$ has energy dimension -2.

If the colliding particles are leptons, the cross section for
quarkonium production is given directly by the factorization 
formula (\ref{sig-NRQCD}).  If the colliding particles 
are hadrons, the cross section (\ref{sig-NRQCD})
must be folded with parton distributions
for the partons 1 and 2 in the colliding hadrons.  
In this case, the derivation of the factorization formula requires that
the transverse momentum $p_T$ of the quarkonium be large
compared to $\Lambda_{QCD}$, the scale of nonperturbative effects in QCD.
This restriction to large $p_T$ follows from the 
diagrammatic analysis that underlies the factorization formula \cite{BBL}.
This analysis shows that the dominant contributions to the cross section
can be factored into (a) hard-scattering amplitudes for parton processes 
of the form $1 2 \to c \bar c +3 4 \ldots$, (b) jet-like subdiagrams for the 
incoming partons 1 and 2 and the outgoing partons 3, 4, etc., 
(c) a subdiagram involving a 
$c \bar c$ pair with relative momentum that is small compared to the 
quark mass $m_c$, and (d) a soft part.
The soft part involves soft gluons that couple to the jet-like subdiagrams 
and to the $c \bar c$ subdiagram, but not to the hard-scattering subdiagram.
The effects of soft partons that are exchanged between the various subdiagrams
cancel upon summing over all possible connections of the soft partons.
This cancellation is effective provided that the $c \bar c$ pair has
large transverse momentum relative to the incoming hadrons. 
All the effects of soft partons can then be factored into parton distributions 
associated with the incoming hadrons, fragmentation functions associated
with the outgoing partons produced by the hard scattering, and a factor 
associated with the $c \bar c$ pair that depends on their relative momentum 
$q$.  In this step in the derivation of the factorization formula,
the short-distance scales $p_T$ and $m_c$ are separated from the 
long-distance scale $\Lambda_{QCD}$.
The effects of the short-distance scales appear only
in the hard-scattering amplitudes, and all effects of the  scale 
$\Lambda_{QCD}$ are factored into parton distributions, fragmentation functions,
and $c \bar c$ factors.

The remaining step in the derivation of the factorization formula
involves separating the scale $m_c$ in the hard-scattering amplitude
from the scale $m_c v$ of the relative momentum in the charmonium state.
This separation can be accomplished  by
Taylor-expanding the hard-scattering amplitude in powers of $q$.
A naive Taylor expansion generates ultraviolet divergences in the 
$c \bar c$ factor.  The effective field theory framework provided by NRQCD 
makes this Taylor expansion meaningful.
The long-distance factors generated by  the Taylor expansion 
can be identified with matrix elements of local operators in NRQCD.  
The renormalization framework of NRQCD allows the ultraviolet divergences 
in these matrix elements to be systematically removed.  
The resulting expression for the cross section has the form of 
the NRQCD factorization formula  (\ref{sig-NRQCD}).
The effects of the scale $m_c$ appear only in the short-distance 
coefficients $C_{nm}$.  All effects involving momentum scales 
of order $m_c v$ or smaller have been  
factored into the NRQCD matrix elements
$\langle {\cal O}_{mn}^{H(\lambda)} \rangle $.
 
The breakdown of the factorization formula at small $p_T$ when the
colliding particles are hadrons can be attributed to ``higher-twist'' 
processes.  In the factorization formula  (\ref{sig-NRQCD}),
the $c$ and $\bar c$ that form charmonium are assumed to be produced 
by a hard-scattering collision involving a single parton from each of the 
colliding hadrons.  Higher-twist processes can involve more than one 
parton from a single hadron.  An example of a higher-twist process for 
charmonium production is one in which the $c$ and $\bar c$ that bind 
to form charmonium are produced by a quantum fluctuation of the hadron 
into a state containing a $c \bar c$ pair \cite{brodsky}.  
The transition of the $c \bar c$ to a 
charmonium state can then be induced by the collision with the other 
hadron.  The short-distance part of this process is the 
fluctuation of the hadron into a state containing the $c \bar c$ pair.
The cross section for the scattering of the $c \bar c$ pair
from the other hadron involves long-distance effects that cannot
be expressed in terms of matrix elements of local NRQCD operators.

\section{Short-distance Coefficients}

The short-distance coefficients $C_{mn}$ in (\ref{sig-NRQCD})
can be determined by matching perturbative calculations in full QCD 
with the corresponding perturbative calculations in NRQCD. 
In most previous calculations, the coefficients were determined 
by matching cross sections for producing $c \bar c$ pairs with 
definite total angular momentum.  
However, $c \bar c$ states with different angular-momentum quantum numbers 
can have a nonzero overlap with the same final state 
$|H + S \rangle$.  The resulting interference terms cannot always
be obtained by matching cross sections.  Below we present a
matching prescription which is sufficiently general to 
provide these interference terms.

\subsection{General matching prescription}

Let $c \bar c({\bf q},\xi,\eta)$ represent a state that consists of a $c$ and a 
$\bar c$ that have total momentum $P$, 3-momenta $\pm {\bf q}$ in the 
$c \bar c$ rest frame, and spin and color states specified by the spinors
$\xi$ and $\eta$.  These spinors are 2-component Pauli spinors with a 
color index.
Using the abbreviated notation 
$c \bar c \equiv c \bar c({\bf q},\xi,\eta)$ and
$c \bar c' \equiv c \bar c({\bf q}',\xi',\eta')$, the matching condition is
\begin{eqnarray}
\sum_X (2 \pi)^4 \delta^4(k_1 + k_2 - P - k_X) \;
	{\cal T}_{1 2 \to c \bar c' + X}^* \; 
	{\cal T}_{1 2 \to c \bar c + X} & &
\nonumber \\
& & \hspace{-2in}
\;=\; \sum_{mn} C_{mn}(P,k_1,k_2) \;
	\langle \chi^\dagger {{\cal K}'}_m^\dagger \psi \; 
	{\cal P}_{c \bar c',c \bar c} \;
	\psi^\dagger {\cal K}_n \chi \rangle .	
\label{match}
\end{eqnarray}
The operator ${\cal P}_{c \bar c',c \bar c}$ in the matrix 
element in (\ref{match}) is defined by 
\begin{equation}
{\cal P}_{c \bar c',c \bar c} 
\;=\; \sum_S | c({\bf q}',\xi') \bar c(-{\bf q}',\eta') + S \rangle 
	\langle c({\bf q},\xi) \bar c(-{\bf q},\eta) + S | .	
\label{P_cc}
\end{equation}
The sum is over soft parton states whose total energy is  less than the 
ultraviolet cutoff $\Lambda$ of NRQCD. 
The normalization of the 
$c \bar c$ states in (\ref{P_cc}) must coincide with those in the 
T-matrix elements in (\ref{sig-M}).  The standard relativistic 
normalization is
\begin{equation}
\langle c({\bf q}_1',\xi') \bar c({\bf q}_2',\eta')
	 | c({\bf q_1},\xi) \bar c({\bf q_2},\eta)  \rangle 
\;=\; 4 E_{q_1} E_{q_2} \; (2 \pi)^6  \delta^3({\bf q}_1 - {\bf q}_1')
	\delta^3({\bf q}_2 - {\bf q}_2') \; 
	\xi^\dagger \xi' {\eta'}^\dagger \eta ,
\label{cc-norm}
\end{equation}
where the spinors are normalized so that 
$\xi^\dagger \xi = \eta^\dagger \eta = 1$, 
and similarly for $\xi'$ and $\eta'$.
In expressions like $\xi^\dagger \xi'$,
both the spin and color indices are contracted.  
Note that with the normalization (\ref{cc-norm}), the operator
$P_{cc',cc}$ in (\ref{P_cc}) has dimension -4. The difference in the 
dimensions of the operators $P_H$ and $P_{cc',cc}$ matches the difference in
the dimensions of the T-matrix elements in (\ref{sig-M}) and (\ref{match}).

To carry out the matching procedure, the left side of (\ref{match}) is
calculated using perturbation theory in full QCD, and then expanded as
a Taylor series in ${\bf q}$ and ${\bf q}'$. The matrix elements on the right 
side of (\ref{match}) are 
calculated using perturbation theory in NRQCD, and then expanded as
Taylor series in ${\bf q}$ and ${\bf q}'$.  The short-distance coefficients
$C_{mn}$ are obtained by matching the terms in these Taylor expansions
order by order in $\alpha_s$.

\subsection{Matching for $q \bar q \to c \bar c$}

We first illustrate the matching procedure by applying it to the process
$q \bar q \to c \bar c$. The T-matrix element in Feynman gauge is 
\begin{equation}
{\cal T}_{1 2 \to c \bar c} 
\;=\; g^2 {1 \over P^2} \; {\bar v}(k_2) \gamma_\mu T^a u(k_1) \;
	\bar u(p) \gamma^\mu T^a v({\bar p}) .
\label{T-qqbar}
\end{equation}
Dirac and color indices are implicit in the spinors.  The 4-momenta of the 
outgoing $c$ and $\bar c$ can be expressed as $p = {1 \over 2} P + L {\bf q}$ 
and $\bar p = {1 \over 2} P - L {\bf q}$, where $P$ is the total 4-momentum 
of the $c \bar c$ pair, ${\bf q}$ is the relative 3-momentum of the $c$ 
in the center-of-momentum (CM) frame of the pair, and $L^\mu _{\ i}$ is a 
Lorentz boost matrix.  The boost matrix $L^\mu _{\ i}$ transforms a purely 
spacelike 4-vector, such as $(0,{\bf q})$,  from the 
CM frame where the components of $P$ are $(2 E_q,{\bf 0})$ to the frame 
in which its components are $P^\mu$.  The matrix is given explicitly in
Appendix A.  The expressions for the Dirac spinors $u(p)$ and $v(\bar p)$
in terms of Pauli spinors $\xi$ and $\eta$ are also given in Appendix A.
Identities in Appendix A can be used to expand 
${\bar u}(p) \gamma^\mu T^a v({\bar p})$ as a Taylor series in ${\bf q}$.
To linear order in ${\bf q}$, the T-matrix element (\ref{T-qqbar}) is
\begin{equation}
{\cal T}_{1 2 \to c \bar c} 
\;=\; {g^2  \over 2 m_c} \; {\bar v}(k_2) \gamma_\mu T^a u(k_1) \;
	L^\mu_{\ i} \; \xi^\dagger \sigma^i T^a \eta .
\label{T-qqbar0}
\end{equation}
The components of the boost matrix to linear order in ${\bf q}$ are
\begin{mathletters}
\begin{eqnarray}
L^0_{\ i} &\;=\;& {1 \over 2 m_c} P^i,
\\
L^j_{\ i} &\;=\;& \delta^{ji} - {\hat P}^j {\hat P}^i
	\;+\; {E_P \over 2 m_c} {\hat P}^j {\hat P}^i,
\label{boost}
\end{eqnarray}
\end{mathletters}
where $E_P = \sqrt{4 m_c^2 + {\bf P}^2}$.

In order to carry out the matching procedure, we first calculate the 
left side of (\ref{match}).  In this case, there is no sum over $X$.
Multiplying (\ref{T-qqbar0}) by the complex conjugate of 
${\cal T}_{1 2 \to c \bar c'} $ and
averaging over initial spins and colors, we obtain
\begin{eqnarray}
(2 \pi)^4 \delta^4(k_1 + k_2 - P) \; \overline{\sum} \; 
	{\cal T}_{1 2 \to c \bar c'}^* \; 
	{\cal T}_{1 2 \to c \bar c} &&
\nonumber \\
&& \hspace{-2in}
\;=\;  (2 \pi)^4 \delta^4(k_1 + k_2 - P) \; {4 \pi^2 \alpha_s^2 \over 9} 
	\left[ \delta^{ji} - {\hat n}^j {\hat n}^i \right]
	{\eta'}^\dagger \sigma^j T^a \xi' \xi^\dagger \sigma^i T^a \eta .	
\label{TT-qqbar}
\end{eqnarray}
We have simplified the expression by using the identities 
(\ref{id:PL}) and (\ref{id:gL}) in Appendix A. We have also used the fact that
$(k_1 - k_2)_\mu L^\mu_{\ i} = - \sqrt{P^2} \, {\hat n}^i$, where $\bf {\hat n}$
is a unit vector, which follows from (\ref{id:LL}).
In a frame in which $\bf P$, $\bf k_1$, and $\bf k_2$ are collinear,
they are also collinear with $\bf {\hat n}$.
 
We next consider the right side of the matching equation (\ref{match}).  
We must construct an operator of the form (\ref{O-NRQCD})
whose vacuum matrix element, when calculated to leading order in $\alpha_s$,
reproduces the spinor factor in (\ref{TT-qqbar}) upon
expanding to linear order in ${\bf q}$ and ${\bf q}'$. One can see
by inspection that the appropriate operator is
$\chi^\dagger \sigma^j T^a \psi {\cal P}_{c \bar c',c \bar c} 
	\psi^\dagger \sigma^i T^a \chi$.  
At leading order in $\alpha_s$, only the $c \bar c$ states
in the projection operator $P_{cc',cc}$ contribute:
\begin{equation}
\langle \chi^\dagger \sigma^j T^a \psi \; 
	{\cal P}_{c \bar c',c \bar c} \;
	\psi^\dagger \sigma^i T^a \chi \rangle
\;=\; \langle 0 | \chi^\dagger \sigma^j T^a \psi 
	| c \bar c ({\bf q}',\xi',\eta') \rangle \; 
	\langle c \bar c ({\bf q},\xi,\eta) |
	\psi^\dagger \sigma^i T^a \chi | 0 \rangle .
\label{M-cc}
\end{equation}
The vacuum-to-$c \bar c$ matrix element reduces at leading order in 
$\alpha_s$ to
\begin{equation}
\langle c \bar c ({\bf q},\xi,\eta) |
	\psi^\dagger \sigma^i T^a \chi | 0 \rangle 
\;=\; 2 E_q \; \xi^\dagger \sigma^i T^a \eta ,
\end{equation}
where the factor of $2 E_q$ arises from the relativistic 
normalization of states.  Expanding to linear order in ${\bf q}$
 and ${\bf q}'$, the matrix element (\ref{M-cc}) reduces to 
\begin{equation}
\langle \chi^\dagger \sigma^j T^a \psi \; 
	{\cal P}_{c \bar c',c \bar c} \;
	\psi^\dagger \sigma^i T^a \chi \rangle
\;=\; 4 m_c^2 \;  {\eta'}^\dagger \sigma^j T^a \xi' 
	\xi^\dagger \sigma^i T^a \eta ,
\label{M-qqbar}
\end{equation}
Comparing (\ref{TT-qqbar}) and (\ref{M-qqbar}),
we can read off the short-distance coefficient for the matrix element:
\begin{equation}
C^{ji} \; = \; (2 \pi)^4 \delta^4 (k_1 + k_2 - P) \,
	{\pi^2 \alpha_s^2 \over 9 m_c^2} \,
	\left[ \delta^{ji} - {\hat n}^j {\hat n}^i \right] .
\label{C-qqbar}
\end{equation}

Having determined the short-distance coefficient, we can insert it into 
(\ref{sig-NRQCD}) to obtain an expression for the differential cross section 
for producing a meson $H$ with helicity $\lambda$ via the parton process 
$q \bar q \to c \bar c$:  
\begin{eqnarray}
d \sigma(q \bar q \to H(P,\lambda)) 
&\;=\;& {1 \over 4 E_1 E_2 v_{12}} \; {d^3P \over (2 \pi)^3 2 E_P} \; 
	(2 \pi)^4 \delta^4(k_1 + k_2 - P) \;
\nonumber \\
&&  \hspace{.5in}
\times {\pi^2 \alpha_s^2 \over 9 m_c^2} 
	\left[ \delta^{ji} - {\hat n}^j {\hat n}^i \right] 
	\langle \chi^\dagger \sigma^j T^a \psi \; 
		{\cal P}_{H(\lambda)} \;
		\psi^\dagger \sigma^i T^a \chi \rangle .
\label{dsig-qqbar}
\end{eqnarray}
Using the delta function to integrate over the phase space of $P$,
we find that the cross section is
\begin{equation}
\sigma(q \bar q \to H(\lambda)) 
\;=\;  \delta(s - 4 m_c^2) \;
	{\pi^3 \alpha_s^2 \over 36 m_c^4} 
	\left[ \delta^{ji} - {\hat z}^j {\hat z}^i \right] 
	\langle \chi^\dagger \sigma^j T^a \psi \; 
		{\cal P}_{H(\lambda)} \;
		\psi^\dagger \sigma^i T^a \chi \rangle ,
\label{sig-qqbar}
\end{equation}
where $s = (k_1 + k_2)^2$ 
and we have assumed that the 3-momentum of the colliding partons is along 
the $z$-axis.  Note that the expression (\ref{sig-qqbar}) is dimensionally 
correct.  Each of the quark fields has dimension $\mbox{${3 \over 2}$}$
and the projection operator has dimension -2, so the dimension of the 
matrix element is 4.  
The expression (\ref{sig-qqbar}) gives a
contribution to the cross section for every quarkonium state $H$.

\subsection{Matching for $g g \to c \bar c$}

 We next illustrate the matching procedure by applying it to the process
 $gg \to c \bar c$. The T-matrix element can be decomposed into 3 
 independent color structures:
\begin{equation}
  {\cal{T}}_{gg\to c\bar{c}}
    = -g^2 \epsilon^a_\mu (k_1) \epsilon^b_\nu (k_2)
 \left( 
   {1 \over 6} \delta^{ab} S^{\mu\nu} +
   {1 \over 2 } d^{abc} D ^{\mu\nu c} + 
   {i \over 2 } f^{abc} F^{\mu\nu c} 
 \right).
\label{mgg}
\end{equation}
The Dirac factor for the  $\delta^{ab}$ term is 
\begin{equation}
S^{\mu\nu}= \bar u(p)
\left[ {\gamma^\mu ( \not \! p \; - \not \! k_1 + m_c) \gamma^\nu 
	\over 2 p \cdot k_1 }
\;+\; {\gamma^\nu ( \not \! p \; - \not \! k_2 + m_c) \gamma^\mu 
	\over 2 p \cdot k_2 } \right] 
 v(\bar p) .
\label{sterm}
\end{equation}
The term $D^{\mu\nu c}$ in (\ref{mgg}) differs from (\ref{sterm}) only by 
inserting the color matrix $T^c$ between the spinors.
The Dirac factor for the $f^{abc}$ term in Feynman gauge is 
\begin{eqnarray}
F^{\mu \nu c} &\;=\;& \bar u(p)
 \Bigg[ 
        {\gamma^\mu ( \not \! p \; - \not \! k_1 + m_c) \gamma^\nu 
           \over 2 p \cdot k_1 }
  \;-\; {\gamma^\nu ( \not \! p \; - \not \! k_2 + m_c) \gamma^\mu 
	   \over 2 p \cdot k_2 }  \nonumber \\
&& \hspace{1in}
\;-\; 2
{ (P +k_2)^\mu \gamma^\nu - (P + k_1)^\nu \gamma^\mu  
	+ g^{\mu \nu} (\not \! k_1 - \not \! k_2 )
	\over P^2 }       
\Bigg] T^c v(\bar p) .
\label{fterm}
\end{eqnarray}
Using the identities in Appendix A, we expand (\ref{sterm}) and 
(\ref{fterm}) to linear order in ${\bf q}$: 
\begin{eqnarray}
S^{\mu \nu} 
&\;=\;& {i \over 2 m_c^2} \, \epsilon^{\mu \nu \lambda \rho} 
	(k_1 - k_2)_\lambda P_\rho 
	\; \xi^\dagger \eta
\nonumber \\
&&
\;+\;  \Bigg[ 
{1 \over m_c^3} (k_1 \cdot L)_n 
	\left( (k_1-k_2)^\mu L^\nu_{\ j} + (k_1-k_2)^\nu L^\mu_{\ j} \right)
\;-\; {1 \over m_c^3} (k_1 \cdot L)_j 
	( P^\mu L^\nu_{\ n} - P^\nu L^\mu_{\ n})
\nonumber \\
&& \hspace{.5in}
\;-\; {2 \over m_c^3} g^{\mu \nu} (k_1 \cdot L)_n (k_1 \cdot L)_j
\;+\; {2 \over m_c} 
	\left( L^\mu_{\ n} L^\nu_{\ j} + L^\nu_{\ n} L^\mu_{\ j} \right)
\Bigg] 
	q^n \; \xi^\dagger \sigma^j \eta ,
\label{stermq}
\\
F^{\mu \nu c} 
&\;=\;&  - {1 \over m} \left( k_1^\mu L^\nu_{\ j} - k_2^\nu L^\mu_{\ j} \right)
	\xi^\dagger \sigma^j T^a \eta
\;-\; {i \over 2 m_c^4} \,
\epsilon^{\mu \nu \lambda \rho} (k_1 - k_2)_\lambda P_\rho 
	(k_1 \cdot L)_n \; q^n \; \xi^\dagger T^c \eta .
\label{ftermq}
\end{eqnarray}
The first term in $F^{\mu \nu c}$ vanishes when contracted with the 
polarization vectors $\epsilon^a_\mu (k_1)$ and $\epsilon^b_\nu (k_2)$
in (\ref{mgg}).
Multiplying (\ref{mgg}) by the complex conjugate of 
${\cal T}_{gg \to c\bar{c}'}$
and averaging over the spins and colors of the initial gluons, we obtain
\begin{eqnarray}
\overline{\sum} \; {\cal T}_{g g \to c \bar c'}^* \; 
	{\cal T}_{g g \to c \bar c}
&\;=\;&  {\pi^2 \alpha_s^2 \over 9} 
\Bigg\{ {\eta'}^\dagger \xi' \xi^\dagger \eta 	
\;+\; {15 \over 8} \; {\eta'}^\dagger T^a \xi' \xi^\dagger T^a \eta
\nonumber \\   
&& 
\;+\; {1 \over m_c^2} \;  
\left[ ( \delta^{m n} - {\hat n}^m {\hat n}^n )  	
	( \delta^{i j} - {\hat n}^i {\hat n}^j )  	
\;+\; ( \delta^{m j} - {\hat n}^m {\hat n}^j )  	
        ( \delta^{n i} - {\hat n}^n {\hat n}^i ) 
\right.
\nonumber \\   
&& \hspace{1in}
\left.
\;+\; \delta^{m i} \delta^{n j} 	
\;-\; ( \delta^{m i} - {\hat n}^m {\hat n}^i )  	
        ( \delta^{n j} - {\hat n}^n {\hat n}^j )  \right]  	
\nonumber \\   
&&  \hspace{1in}
\times \; {q'}^m q^n
\left( {\eta'}^\dagger \sigma^i \xi' \xi^\dagger \sigma^j \eta 
	\;+\; {15 \over 8} \; 
	{\eta'}^\dagger \sigma^i T^a \xi' \xi^\dagger \sigma^j T^a \eta 
	\right)
\nonumber \\   
&& 
\;+\; {27 \over 8m_c^2} \; {\hat n}^m {\hat n}^n \;
	{q'}^m q^n \; {\eta'}^\dagger T^a \xi' \xi^\dagger T^a \eta
\Bigg\} .
\label{TT-gg}
\end{eqnarray}
At leading order in $\alpha_s$ and to linear order  
in ${\bf q}$ and ${\bf q}'$, the spinor factors on the right side of 
(\ref{TT-gg}) can be identified with  the following 
matrix elements:
\begin{mathletters}
\begin{eqnarray}
\langle \chi^\dagger \psi \, {\cal P}_{c\bar{c}',c\bar{c}} \,
	\psi^\dagger \chi \rangle
&\;=\;& 4 m_c^2 \; \eta'^\dagger \xi' \xi^\dagger \eta  ,
\\
\langle \chi^\dagger T^a \psi \, {\cal P}_{c\bar{c}',c\bar{c}} \,
	\psi^\dagger T^a \chi \rangle
&\;=\;& 4 m_c^2 \; \eta'^\dagger T^a \xi' \xi^\dagger T^a \eta  ,
\\
\langle \chi^\dagger (-\mbox{$\frac{i}{2}$} \tensor{D}^m) T^a \psi \, 
	{\cal P}_{c\bar{c}',c\bar{c}} \,
        \psi^\dagger (-\mbox{$\frac{i}{2}$} \tensor{D}^n) T^a \chi \rangle
&\;=\;& 4 m_c^2 \; {q'}^m q^n \; \eta'^\dagger T^a \xi' \xi^\dagger T^a \eta  ,
\\
\langle \chi^\dagger (-\mbox{$\frac{i}{2}$} \tensor{D}^m) \sigma^i T^a \psi \, 
	{\cal P}_{c\bar{c}',c\bar{c}} \,
	\psi^\dagger (-\mbox{$\frac{i}{2}$} \tensor{D}^n) \sigma^j T^a \chi \rangle
&\;=\;& 4 m_c^2 \; {q'}^m q^n \; \eta'^\dagger \sigma^i T^a \xi'
	\xi^\dagger \sigma^j T^a \eta ,
\end{eqnarray}
\end{mathletters}
where $\psi^\dagger \tensor{\bf D} \chi 
	= \psi^\dagger {\bf D} \chi  - ({\bf D} \psi)^\dagger \chi$.
 
Matching terms on the left side and right side of (\ref{match}), 
we determine the short-distance coefficient $C_{mn}$ for each of the 
matrix elements. Inserting these coefficients into
(\ref{sig-NRQCD}), we obtain the differential cross section for production of 
$H(\lambda)$ from the parton process $gg\to c \bar{c}$. The final result,
after integrating over phase space,  is
\begin{eqnarray}
\sigma( g g \to H(\lambda) ) 
&\;=\;& \delta ( s - 4 m_c^2 ) {\pi^3 \alpha_s^2 \over 144 m_c^4}
\Bigg\{
\langle \chi^\dagger \psi \, {\cal P}_{H(\lambda)} \, \psi^\dagger \chi \rangle
\;+\; {15 \over 8} \; \langle \chi^\dagger T^a \psi \, {\cal P}_{H(\lambda)} \,
	\psi^\dagger T^a \chi \rangle
\nonumber \\   
&& \hspace{-1in}
\;+\; {1 \over m_c^2}
\left[ ( \delta^{m n} - {\hat z}^m {\hat z}^n )  	
        ( \delta^{i j} - {\hat z}^i {\hat z}^j ) 
\;+\; ( \delta^{m j} - {\hat z}^m {\hat z}^j )  	
        ( \delta^{n i} - {\hat z}^n {\hat z}^i ) 
\;+\; \delta^{m i} \delta^{n j} 
\right.
\nonumber \\     
&& \hspace{-.5in}
\left.
\;-\; ( \delta^{m i} - {\hat z}^m {\hat z}^i )  	
        ( \delta^{n j} - {\hat z}^n {\hat z}^j ) 
\right]
\left( \langle \chi^\dagger (-\mbox{$\frac{i}{2}$} \tensor{D}^m) \sigma^i \psi \, 
	{\cal P}_{H(\lambda)} \,
	\psi^\dagger (-\mbox{$\frac{i}{2}$} \tensor{D}^n) \sigma^j \chi \rangle
\right.
\nonumber \\   
&& \hspace{1in}
\left.
\;+\; {15 \over 8} \;
	\langle \chi^\dagger (-\mbox{$\frac{i}{2}$} \tensor{D}^m) \sigma^i T^a \psi \, 
	{\cal P}_{H(\lambda)} \,
	\psi^\dagger (-\mbox{$\frac{i}{2}$} \tensor{D}^n) \sigma^j T^a \chi \rangle \right)
\nonumber \\   
&& \hspace{-1in}
\;+\; {27 \over 8 m_c^2} \;  {\hat z}^m {\hat z}^n \;
	\langle \chi^\dagger (-\mbox{$\frac{i}{2}$} \tensor{D}^m) T^a \psi \, {\cal P}_{H(\lambda)} \,
	\psi^\dagger (-\mbox{$\frac{i}{2}$} \tensor{D}^n) T^a \chi \rangle\Bigg\} .  	
\label{sig-gg}
\end{eqnarray}
The expression (\ref{sig-gg})
for the cross section applies equally well to 
any quarkonium state $H$.  The relative importance of the various terms
for a given state $H$ depends on the magnitude of the matrix elements.

\section{Reducing the Matrix Elements}

The cross sections (\ref{sig-qqbar}) and (\ref{sig-gg}) hold for any
quarkonium state $H$, with the dependence on $H$
appearing only in the matrix elements.  
The matrix elements can be simplified by using rotational symmetry and the 
approximate heavy-quark spin symmetry of NRQCD.  
Their magnitudes can be estimated using velocity-scaling rules for NRQCD
matrix elements.  
In this section, we apply these methods to simplify the matrix
elements that appear in inclusive cross sections for the 
production of $J/\psi$.
The extension to other quarkonium states will be presented elsewhere
\cite{bc:next}.

\subsection{Rotational symmetry}

Under rotations, each of the matrix elements in (\ref{sig-qqbar}) and 
(\ref{sig-gg}) transforms as a component of a tensor.  The indices of the
tensor include vector indices, such as $i$, $j$, $m$, and $n$, and the 
helicity $\lambda$ in the projection operator ${\cal P}_{H(\lambda)}$. 
The helicity $\lambda$ appears both in the ket and in the bra in the 
expression for the projection operator given in (\ref{P_H}).
If the meson has total angular momentum $J$, then
${\cal P}_{H(\lambda)}$ transforms like the $(\lambda,-\lambda)$ component
of the direct product representation $\underline{J} \otimes \underline{J}$.
Thus the entire operator transforms as an element of the representation 
$\underline{J} \otimes \underline{J} \otimes \underline{1} 
	\otimes \underline{1} \otimes ...$, where there is a 
$\underline{1}$ for each vector index of the operator.
The vacuum matrix element of the operator can be expressed 
in terms of $N$ independent matrix elements, where $N$ is the number of times
the trivial representation $\underline{0}$ appears 
in the decomposition  of the direct product 
$\underline{J} \otimes \underline{J} \otimes \underline{1} 
	\otimes \underline{1} \otimes ...$
into irreducible representations.

In the case of the $J/\psi$, the independent  matrix elements can be 
determined by elementary tensor analysis.  The $\psi$ has total angular
momentum $J=1$, and the helicity label $\lambda$ is just a vector index 
in a spherical basis.
The spherical basis and the Cartesian basis are related by a unitary 
transformation with matrix
\begin{equation}
U_{\lambda i}  \;=\;
\left( \begin{array}{ccc} 
	-1/\sqrt{2} & -i/\sqrt{2} &  \ \ 0 \ \ \\ 
	     0      &      0      &  \ \ 1 \ \ \\
	 1/\sqrt{2} & -i/\sqrt{2} &  \ \ 0 \ \ \\ 
	\end{array} \right)_{\lambda i} .
\label{U-matrix}
\end{equation}
The helicity labels $\lambda$
in the ket and in the bra in (\ref{P_H}) are related to 
the corresponding Cartesian indices $i$ and $j$ by the transformation matrices
$U_{\lambda i}$ and $U^\dagger_{j \lambda}$, respectively.
Since there are no momentum vectors that the matrix 
element can depend on, it must be expressible in terms of the 
invariant tensors $\delta^{mn}$ and $\epsilon^{lmn}$.  The number of 
independent matrix elements is the number of invariant tensors that can
be formed out of the available indices.

We first consider operators with no vector indices, such as
$\langle \chi^\dagger \psi {\cal P}_{\psi(\lambda)} \psi^\dagger \chi \rangle$.  
The only indices that
are available are $\lambda$ and $\lambda$ from the projection operator
${\cal P}_{\psi(\lambda)}$.  The only invariant tensor is 
$\delta_{\lambda \lambda}=1$.  Therefore, we have
\begin{equation}
\langle \chi^\dagger \psi \, {\cal P}_{\psi(\lambda)} \, 
	\psi^\dagger \chi \rangle
\;=\; {1 \over 3} \langle \chi^\dagger  \psi \, 
	{\cal P}_\psi \, 
	\psi^\dagger  \chi \rangle ,
\end{equation}
where ${\cal P}_\psi = \sum_\lambda {\cal P}_{\psi(\lambda)}$.
Standard production matrix elements ${\cal O}^\psi_1({}^{2S+1}L_J)$
and ${\cal O}^\psi_8({}^{2S+1}L_J)$ were defined in Ref. \cite{BBL}
using a projection operator analogous to (\ref{P_H}), but constructed 
out of states with the standard nonrelativistic normalization.
As discussed in Appendix B, the two projection operators differ at leading
order simply by a normalization factor of $4 m_c$.  Thus, we can write
\begin{mathletters}
\label{RS:1S}
\begin{eqnarray}
\langle \chi^\dagger \psi \, {\cal P}_{\psi(\lambda)} \, 
	\psi^\dagger \chi \rangle
&\;=\;& {4 \over 3} \; m_c \; \langle {\cal O}^\psi_1({}^1S_0) \rangle ,
\\
\langle \chi^\dagger T^a \psi \, {\cal P}_{\psi(\lambda)} \, 
	\psi^\dagger T^a \chi \rangle
&\;=\;& {4 \over 3} \; m_c \; \langle {\cal O}^\psi_8({}^1S_0) \rangle ,
\label{RS:81S}
\end{eqnarray}
\end{mathletters}
where the matrix elements on the right sides are defined in Appendix B.

We next consider matrix elements with two vector indices, such as 
$\langle \chi^\dagger \sigma^i \psi {\cal P}_{\psi(\lambda)} 
	\psi^\dagger \sigma^j \chi \rangle$.  
Rotational symmetry implies that such a matrix element can  
be expressed in terms of 3 independent tensors: 
$\delta^{ij}$, $U_{\lambda i} U^\dagger_{j \lambda}$,
and $U_{\lambda j} U^\dagger_{i \lambda}$.
Thus there are three independent matrix elements.
If we sum over the helicities $\lambda$, the only possible tensor is 
$\delta^{ij}$, so there is only one possible matrix element.
Thus, up to corrections of relative order $v^2$, we have
\begin{mathletters}
\begin{eqnarray}
\sum_\lambda
\langle \chi^\dagger \sigma^i \psi \, {\cal P}_{\psi(\lambda)} \, 
	\psi^\dagger \sigma^j \chi \rangle
&\;=\;& {4 \over 3} \; \delta^{ij} \; m_c \; 
	\langle {\cal O}^\psi_1({}^3S_1) \rangle ,
\\
\sum_\lambda
\langle \chi^\dagger \sigma^i T^a \psi \, {\cal P}_{\psi(\lambda)} \, 
	\psi^\dagger \sigma^j T^a \chi \rangle
&\;=\;& {4 \over 3} \; \delta^{ij} \; m_c \; 
	\langle {\cal O}^\psi_8({}^3S_1) \rangle ,
\\
\sum_\lambda
\langle \chi^\dagger (-\mbox{$\frac{i}{2}$} \tensor{D}^i) \psi \, 
	{\cal P}_{\psi(\lambda)} \, 
	\psi^\dagger (-\mbox{$\frac{i}{2}$} \tensor{D}^j) \chi \rangle
&\;=\;&  {4 \over 3} \; \delta^{ij} \; m_c  \; 
	\langle {\cal O}^\psi_1({}^1P_1) \rangle ,
\\
\sum_\lambda
\langle \chi^\dagger (-\mbox{$\frac{i}{2}$} \tensor{D}^i) T^a \psi \, 
	{\cal P}_{\psi(\lambda)} \, 
	\psi^\dagger (-\mbox{$\frac{i}{2}$} \tensor{D}^j) T^a \chi \rangle
&\;=\;&  {4 \over 3} \; \delta^{ij} \; m_c \; 
	\langle {\cal O}^\psi_8({}^1P_1) \rangle ,
\label{RS:81P}
\end{eqnarray}
\end{mathletters}
where the matrix elements on the right sides are defined in Appendix B.
In the cross section (\ref{sig-gg}), we also require the matrix element 
$\langle \chi^\dagger (-\mbox{$\frac{i}{2}$} \tensor{D}^j) T^a \psi 
	{\cal P}_{\psi(\lambda)}
	\psi^\dagger (-\mbox{$\frac{i}{2}$} \tensor{D}^j) T^a \chi \rangle$ 
contracted with ${\hat z}^i {\hat z}^j$.  This can be expressed in terms of 
$\langle {\cal O}^\psi_8({}^1P_1) \rangle$ and one additional matrix element:
\begin{eqnarray}
{\hat z}^i {\hat z}^j \; \langle \chi^\dagger 
	(-\mbox{$\frac{i}{2}$} \tensor{D}^i) T^a \psi \, 
	{\cal P}_{\psi(\lambda)} \, 
	\psi^\dagger (-\mbox{$\frac{i}{2}$} \tensor{D}^j) T^a \chi \rangle
&\;=\;&  {2(1 - \delta_{\lambda 0}) \over 3} \; m_c \;
	\langle {\cal O}^\psi_8({}^1P_1) \rangle
\nonumber \\
&& \hspace{-1in}
\;+\; {3 \delta_{\lambda 0} - 1 \over 2}\;
\langle \chi^\dagger (-\mbox{$\frac{i}{2}$} \tensor{D}^3) T^a \psi \, 
	{\cal P}_{\psi(0)} \, 
	\psi^\dagger (-\mbox{$\frac{i}{2}$} \tensor{D}^3) T^a \chi \rangle .
\label{SS:1P1}
\end{eqnarray}
Upon summing over helicities, we recover (\ref{RS:81P}).

Matrix elements with 4 vector indices, such as 
$\langle \chi^\dagger \sigma^i (-\mbox{$\frac{i}{2}$} \tensor{D}^m) \psi 
	{\cal P}_{H(\lambda)} 
	\psi^\dagger \sigma^j (-\mbox{$\frac{i}{2}$} \tensor{D}^n) \chi \rangle$,
can be reduced by rotational symmetry to 15 independent matrix elements.
Upon summing over helicities, the number of independent matrix elements 
is reduced to 3.

\subsection{Vacuum-saturation approximation}

The vacuum matrix element
$\langle \chi^\dagger \sigma^i \psi {\cal P}_{\psi(\lambda)} 
	\psi^\dagger \sigma^j \chi \rangle$
can be simplified by using the {\it vacuum-saturation approximation}.
In this approximation, only the NRQCD vacuum is retained in the sum over 
soft states in the projection operator (\ref{P_H}).  The matrix element
then reduces to
\begin{equation}
\langle \chi^\dagger \sigma^i \psi \, {\cal P}_{\psi(\lambda)} \, 
	\psi^\dagger \sigma^j \chi \rangle
\;\approx\; 
\langle 0 | \chi^\dagger \sigma^i \psi | 
	\psi({\bf P}=0,\lambda) \rangle \, \langle \psi({\bf P}=0,\lambda) | 
	\psi^\dagger \sigma^j \chi | 0 \rangle.
\label{VSA-1}
\end{equation}
As pointed out in ref. \cite{BBL}, the  vacuum-saturation approximation 
is a controlled approximation in the case of heavy quarkonium,
with a relative error of order $v^4$.  The reason for this is that 
the matrix element involves a transition from a
color-singlet $c \bar c$ state to  states of the form 
$| \psi(\lambda) + S \rangle$, where $S$ is a soft state.
Since the $\psi$ is a color singlet,
the state $S$ must also be a color singlet. Suppose it is not the 
NRQCD vacuum.   Since a single chromoelectric dipole transition will 
produce a color-octet state, at least two chromoelectric 
dipole transitions are required to produce the state $S$.  The amplitude
is suppressed by a power of $v$ for each such transition.
There is another power of $v^2$ from the complex conjugate amplitude,
and this gives an overall suppression factor of $v^4$.

After using the  vacuum-saturation approximation, the matrix element
(\ref{VSA-1}) can be further simplified by using rotational symmetry.
The vacuum-to-$\psi$ matrix element must have the form
\begin{equation}
\langle \psi({\bf P}=0,\lambda) | \psi^\dagger \sigma^j \chi | 0 \rangle
\;=\; U^\dagger_{j \lambda} \;\sqrt{2 M_\psi} \; \sqrt{3 \over 2 \pi} 
	\overline{R}_\psi .
\label{R_psi}
\end{equation}
The factor of $\sqrt{2 M_\psi}$ comes from the relativistic 
normalization of the charmonium state.
The factor $\sqrt{3/2 \pi}$ has been inserted so that 
$\overline{R}_\psi$ can be interpreted as the  
nonrelativistic radial wavefunction evaluated at the origin.
Inserting (\ref{R_psi}) into (\ref{VSA-1}), it reduces to 
\begin{equation}
\langle \chi^\dagger \sigma^i \psi \, {\cal P}_{\psi(\lambda)} \, 
	\psi^\dagger \sigma^j \chi \rangle
\;=\; U_{\lambda i} U^\dagger_{j \lambda}  \;
	{3 \over \pi} M_\psi | \overline{R}_\psi |^2 
\;+\; O( v^4 m_c |\overline{R}_\psi |^2 ) .
\label{VSA-2}
\end{equation}

\subsection{Heavy-quark spin symmetry}

The most powerful tool at our disposal for simplifying the 
production matrix elements of heavy quarkonium is the 
approximate heavy-quark spin symmetry of NRQCD.
The spin symmetry transformations of the heavy-quark and antiquark fields are 
\begin{mathletters}
\begin{eqnarray}
\psi({\bf x},t) & \to &  V \; \psi({\bf x},t) ,
\\
\chi({\bf x},t) & \to &  W \; \chi({\bf x},t) ,
\end{eqnarray}
\end{mathletters}
where $V$ and $W$ are SU(2) matrices.
Under a rotation, the transformation of the fields is
\begin{mathletters}
\begin{eqnarray}
\psi({\bf x},t) & \to &  V \; \psi(O.{\bf x},t) ,
\\
\chi({\bf x},t) & \to &  V \; \chi(O.{\bf x},t) ,
\end{eqnarray}
\end{mathletters}
where $O$ is the O(3) matrix whose elements are 
\begin{equation}
O^{ij} \;=\;   {1 \over 2} \; {\rm tr}( V^\dagger \sigma^i V \sigma^j ) .
\end{equation}

While rotational symmetry is an exact symmetry of NRQCD, spin symmetry is 
only an approximate symmetry.
NRQCD is equivalent to full QCD in the sense that
the coefficients of the terms in the NRQCD lagrangian can be tuned
so that the effective theory reproduces the quarkonium spectrum and 
low energy quarkonium matrix elements to any desired order in $v$.
Minimal NRQCD reproduces the spectrum and 
matrix elements to a relative 
accuracy of order $v^2$.  This theory has exact heavy-quark spin symmetry.
With the $v^2$-improved NRQCD lagrangian, the accuracy is improved
to $v^4$.  However the $v^2$-improved lagrangian includes a term 
$\psi^\dagger {\bf B} \cdot \mbox{\boldmath $\sigma$} \psi
- \chi^\dagger {\bf B} \cdot \mbox{\boldmath $\sigma$} \chi$
that breaks the spin symmetry.  Thus spin symmetry gives relations
between matrix elements that are accurate up to corrections of 
relative order $v^2$.

The consequences of spin symmetry are particularly simple for 
spin-triplet matrix elements of the $\psi$, such as
$\langle \chi^\dagger \sigma^i \psi {\cal P}_{\psi(\lambda)} 
	\psi^\dagger \sigma^j \chi \rangle$.
Since the $\psi$ is an S-wave state, its helicity $\lambda$ 
is a spin index.  Spin symmetry implies that the helicity label $\lambda$
in the ket of the projection operator ${\cal P}_{\psi(\lambda)}$
must be matched with the vector index $i$, while the $\lambda$ in the 
bra must be matched with the index $j$.  Thus the matrix element must be 
proportional to $U_{\lambda i} U^\dagger_{j \lambda}$.  The 
proportionality constant is obtained by summing over $\lambda$ and 
contracting with $\delta^{ij}$.  We find that, up to corrections of 
relative order $v^2$,
\begin{mathletters}
\begin{eqnarray}
\langle \chi^\dagger \sigma^i \psi \, {\cal P}_{\psi(\lambda)} \, 
	\psi^\dagger \sigma^j \chi \rangle
&\;=\;& {4 \over 3} \; U_{\lambda i} U^\dagger_{j \lambda}  \; 
	m_c \langle {\cal O}^\psi_1({}^3S_1) \rangle.
\label{SS:13S}
\\
\langle \chi^\dagger \sigma^i T^a \psi \, {\cal P}_{\psi(\lambda)} \, 
	\psi^\dagger \sigma^j T^a \chi \rangle
&\;=\;& {4 \over 3} \; U_{\lambda i} U^\dagger_{j \lambda}  \;
	m_c \langle {\cal O}^\psi_8({}^3S_1) \rangle,
\label{SS:83S}
\end{eqnarray}
\end{mathletters}
The relative error of $v^2$ in (\ref{SS:13S}) is larger than the 
relative error of $v^4$ obtained by using the vacuum saturation 
approximation (\ref{VSA-2}). One can easily show that the error
in (\ref{SS:13S})  can be improved to $v^4$ by replacing
$4 m_c$ by $2 M_\psi$.

The matrix element 
$\langle \chi^\dagger \sigma^i (-\mbox{$\frac{i}{2}$} \tensor{D}^m) \psi 
	{\cal P}_{\psi(\lambda)} 
	\psi^\dagger \sigma^j (-\mbox{$\frac{i}{2}$} \tensor{D}^n) \chi \rangle$
can also be reduced to a single matrix element by using spin symmetry. 
This symmetry implies that the tensor structure in the indices 
$\lambda$, $i$, and $j$ must be the same as in (\ref{SS:83S}).
Rotational symmetry then implies that the 
matrix element must also be proportional to 
$\delta^{mn}$.  Thus, up to corrections of relative order $v^2$, we have
\begin{mathletters}
\label{SS:3P}
\begin{eqnarray}
\langle \chi^\dagger \sigma^i (-\mbox{$\frac{i}{2}$} \tensor{D}^m) \psi \, 
	{\cal P}_{\psi(\lambda)} \, 
	\psi^\dagger \sigma^j (-\mbox{$\frac{i}{2}$} \tensor{D}^n) \chi \rangle
\;=\; 4 \; U_{\lambda i} U^\dagger_{j \lambda}  \delta^{mn} \;
	m_c \; \langle {\cal O}^\psi_1({}^3P_0) \rangle,
\label{SS:13P}
\\
\langle \chi^\dagger \sigma^i (-\mbox{$\frac{i}{2}$} \tensor{D}^m) T^a \psi \, 
	{\cal P}_{\psi(\lambda)} \, 
	\psi^\dagger \sigma^j (-\mbox{$\frac{i}{2}$} \tensor{D}^n) T^a \chi \rangle
\;=\; 4 \; U_{\lambda i} U^\dagger_{j \lambda}  \delta^{mn} \; m_c \; 
	\langle {\cal O}^\psi_8({}^3P_0) \rangle .
\label{SS:83P}
\end{eqnarray}
\end{mathletters}

Spin symmetry can also be applied to spin-singlet 
matrix elements of the $\psi$, such as 
$\langle \chi^\dagger (-\mbox{$\frac{i}{2}$} \tensor{D}^i) \psi 
	{\cal P}_{\psi(\lambda)} 
	\psi^\dagger (-\mbox{$\frac{i}{2}$} \tensor{D}^j) \chi \rangle$.  
A spin-singlet operator like 
$\psi^\dagger (-\mbox{$\frac{i}{2}$} \tensor{D}^j) \chi$ 
creates a $c \bar c$ pair in a spin-singlet state.
At leading order in $v^2$, the transition to a final state of the form
$|\psi + S \rangle$ must involve the term
$\psi^\dagger {\bf B} \cdot \mbox{\boldmath $\sigma$} \psi
- \chi^\dagger {\bf B} \cdot \mbox{\boldmath $\sigma$} \chi$
from the $v^2$-improved NRQCD lagrangian.  The resulting constraints
on the matrix element can be deduced using angular momentum theory.
These constraints are simply those that are already provided 
by rotational invariance.
Spin symmetry does give new relations between the production matrix elements 
of the $\psi$ and those of the $\eta_c$, but they are not of great 
practical significance.

\subsection{Velocity-scaling rules}

The relative importance of the NRQCD matrix elements that appear in
cross sections for  $\psi$ production are determined by how they scale 
with the relative velocity $v$ of the charm quark in the meson \cite{BBL}.
When applied to the matrix elements
$\langle {\cal O}_n({}^{2S+1}L_J) \rangle$, the scaling rules are fairly simple.
The operator  ${\cal O}_n({}^{2S+1}L_J)$ creates and annihilates a pointlike $c \bar c$ pair in 
the color state $\underline{n}$
and in the angular momentum state ${}^{2S+1}L_J$.
The scaling of the matrix element is determined by the orbital-angular-momentum 
quantum number $L$ and by the number of chromoelectric and 
chromomagnetic dipole transitions that are required 
for the $c \bar c$ pair to reach the state
$c \bar c (\underline{n},{}^{2S+1}L_J)$ from the dominant Fock state 
of the meson.  If the minimum number of these transitions is $E$ and $M$, 
respectively, the matrix element $\langle {\cal O}_n({}^{2S+1}L_J) \rangle$
scales as $v^{3+2L+2E+4M}$.

The dominant Fock state of the $\psi$ consists of a $c \bar c$ pair in a 
color-singlet ${}^3S_1$ state.  Thus the matrix element 
$\langle {\cal O}_1({}^3S_1) \rangle$ scales as $v^3$.
The $c \bar c$ pair can reach the color-octet states
$c \bar c (\underline{8},{}^3P_J)$, $c \bar c (\underline{8},{}^3S_1)$,
and $c \bar c (\underline{8},{}^1S_0))$  through a single chromoelectric,
a double chromoelectric, and a single chromomagnetic dipole transition,
respectively.  Thus the matrix elements
$\langle {\cal O}_8({}^3P_J) \rangle$, $\langle {\cal O}_8({}^3S_1) \rangle$,
and $\langle {\cal O}_8({}^1S_0) \rangle$ all scale as $v^7$.
All the other matrix elements 
$\langle {\cal O}_n({}^{2S+1}L_J) \rangle$ scale as $v^{11}$ or smaller. 
Those that scale like $v^{11}$ include
$\langle {\cal O}_1({}^1P_1) \rangle$, $\langle {\cal O}_8({}^1P_1) \rangle$,
$\langle {\cal O}_1({}^1S_0) \rangle,$ and 
$\langle {\cal O}_1({}^3P_J) \rangle$.
The matrix element
$\langle \chi^\dagger (-{i \over 2} \tensor{D}^3) T^a \psi \, 
	{\cal P}_{\psi(0)} \,
	\psi^\dagger (-{i \over 2} \tensor{D}^3) T^a \chi \rangle$ 
also scales as $v^{11}$ like $\langle {\cal O}_8({}^1P_1) \rangle$.

\section{Cross Sections for Polarized $\psi$}

In this section, we combine the results of Sections III and IV to 
obtain the cross sections for production of $\psi$ with definite 
helicity via the order-$\alpha_s^2$ parton processes
$q \bar q \to c \bar c$ and $g g \to c \bar c$.  These cross sections
can be folded with parton distributions to obtain 
cross sections for $\psi$ production in hadron-hadron scattering.
The resulting cross sections should not be taken too seriously,
since these parton processes produce a $\psi$ with zero transverse momentum.
The factorization formula (\ref{sig-NRQCD}) is therefore not strictly
applicable, since its derivation requires the transverse momentum of the 
$\psi$ to be large compared to $\Lambda_{QCD}$.  
We choose to ignore this difficulty, since our primary purpose 
is to illustrate the calculation of cross sections for polarized production.

The cross section for $\psi$ production via the parton process 
$q \bar q \to c \bar c$ is obtained by
inserting the matrix element (\ref{SS:83S}) into (\ref{sig-qqbar}):
\begin{equation}
\sigma(q \bar q \to \psi (\lambda)) 
\;=\;  \delta(s - 4 m_c^2) \;
	{\pi^3 \alpha_s^2 \over 27 m_c^3} \;
	\left( 1- \delta_{\lambda 0 }  \right)\;
	\langle O_8^{\psi}(^3S_1) \rangle .
\label{spin-qqbar}
\end{equation}
We have used the fact that the matrix $U$ in (\ref{U-matrix}) satisfies
$\sum_i U_{\lambda i} U^\dagger_{i \lambda} = 1$ and 
$\sum_j U^\dagger_{j \lambda} {\hat z}^j =  \delta_{\lambda 0}$.
The cross section (\ref{spin-qqbar}) implies that
$\psi$'s  produced by this process
are transversely polarized, which could be anticipated from the fact that the 
gluon interaction with the light quark conserves helicity.
Summing over helicities, we obtain
\begin{equation}
\sigma(q \bar q \to \psi ) 
\;=\;  \delta(s - 4 m_c^2) \;
	{2 \pi^3 \alpha_s^2 \over 27 m_c^3} \;
	\langle O_8^{\psi}(^3S_1) \rangle .
\label{nonpol-qqbar}
\end{equation}

The cross section for producing a $\psi$ with helicity $\lambda$ via the 
parton process $g g \to c \bar c$ is obtained by inserting the 
matrix elements 
(\ref{RS:1S}), (\ref{SS:1P1}), and 
(\ref{SS:3P}) into (\ref{sig-gg}):
\begin{eqnarray}
\sigma( g g \to \psi(\lambda) ) 
&\;=\;& \delta ( s - 4 m_c^2 ) \;
{\pi^3 \alpha_s^2 \over 108 m_c^3}
\;\Bigg\{\;
\langle O_1^{ \psi } ( ^1 S_0 ) \rangle 
\;+\; {15 \over 8} \langle O_8^{\psi} ( ^1 S_0 ) \rangle 
\nonumber \\   
&& \hspace{.5in}
\;+\; {3 (3-2\delta_{\lambda 0}) \over m_c^2} 
        \langle O_1^{\psi}(^3 P_0) \rangle 
\;+\; {45 (3-2\delta_{\lambda 0}) \over 8 m_c^2}
	\langle O_8^{\psi}(^3 P_0) \rangle 
\nonumber \\   
&& \hspace{.5in}
\;+\; {27 (1 - \delta_{\lambda 0}) \over 16 m_c^2} 
        \langle O_8^{\psi}( ^1 P_1 ) \rangle 
\nonumber \\   
&& \hspace{.5in}
\;+\; {81 (1 - 3 \delta_{\lambda 0}) \over 64 m_c^3}
	\langle \chi^\dagger (-\mbox{$\frac{i}{2}$} \tensor{D}^3) T^a \psi \, 
	{\cal P}_{\psi(0)} \,
	\psi^\dagger (-\mbox{$\frac{i}{2}$} \tensor{D}^3) T^a \chi \rangle
\;\Bigg\} .  	
\label{spin-gg}
\end{eqnarray}
Summing over helicities, this reduces to 
\begin{eqnarray}
\sigma( g g \to \psi ) 
\;=\;  \delta ( s - 4 m_c^2 ) \;
{\pi^3 \alpha_s^2 \over 36 m_c^3}
\;\Bigg\{ \langle O_1^{ \psi } ( ^1 S_0 ) \rangle 
\;+\; {15 \over 8} \langle O_8^{\psi} ( ^1 S_0 ) \rangle 
\;+\; {7 \over m_c^2} \langle O_1^{\psi}(^3 P_0) \rangle 
\nonumber \\   
\;+\; {105 \over 8 m_c^2}
	\langle O_8^{\psi}(^3 P_0) \rangle 
\;+\; {9 \over 8 m_c^2} \langle O_8^{\psi}( ^1 P_1 ) \rangle 
\Bigg\} .  	
\label{nonpol-gg}
\end{eqnarray}
Note that the contribution from the matrix element
$\langle \chi^\dagger (-{i \over 2} \tensor{D}^3) T^a \psi \, 
	{\cal P}_{\psi(0)} \,
	\psi^\dagger (-{i \over 2} \tensor{D}^3) T^a \chi \rangle$
drops out of the cross section for unpolarized production.
If we retain only those matrix elements that are of leading order in 
$v^2$, the cross section (\ref{spin-gg}) reduces to 
\begin{equation}
\sigma( g g \to \psi(\lambda) ) 
\;\approx\;  \delta ( s - 4 m_c^2 ) \;
{5\pi^3 \alpha_s^2 \over 288 m_c^3}
\;\left\{\;
\langle O_8^{ \psi } ( ^1 S_0 ) \rangle 
\;+\; {3 (3-2\delta_{\lambda 0}) \over m_c^2} \langle O_8^{\psi}(^3 P_0) \rangle 
\;\right\} .  	
\label{spin-gg:LO}
\end{equation}
The $\langle O_8^{ \psi } ( ^1 S_0 ) \rangle$ term gives 
$\psi$'s that are unpolarized,
while the $\langle O_8^{\psi}(^3 P_0) \rangle$ term gives $\psi$'s
with helicity $-1$, 0, and +1 in the proportions $3 \, : \, 1 \, : \, 3$.

The production of  $J/\psi$ from the color-octet parton processes
$q\bar{q} \to c \bar{c}$ and $gg \to c\bar{c}$ has been studied previously by 
several groups \cite{tang-vantinnen,cho-leibovich,fleming-maksymyk}. 
In particular, the spin alignment of the  $\psi$
from these processes has 
been calculated by Tang and V\"{a}ntinnen \cite{tang-vantinnen} 
and by Cho and Leibovich \cite{cho-leibovich}.
We proceed to compare our results with this previous work. 

Tang and V\"{a}ntinnen \cite{tang-vantinnen} have used the 
covariant projection method to calculate the cross sections for polarized
$\psi$ and $\psi'$ in fixed-target hadron-hadron collisions
from the order-$\alpha_s^2$ parton processes.  
Our result (\ref{spin-qqbar}) for the cross section from 
$q\bar{q} \to c \bar{c}$ is in agreement with theirs.  
Our result (\ref{spin-gg:LO}) for the cross section 
from $g g \to c \bar{c}$ agrees with theirs only after summing over helicities.  
We disagree on the helicity dependence of the 
$\langle O_8^{\psi}(^1S_0) \rangle$ term. 
Their result for this term is proportional to 
$1 - \delta_{\lambda 0}$, which implies that the 
$\psi$ is produced with transverse polarization. 
This is clearly incorrect, because a pointlike $c \bar c$ pair in a 
$^1S_0$ state is rotationally invariant in its rest frame.  The $\psi$'s that 
are produced by the binding of the $c \bar c$ pair must therefore be 
unpolarized, as in our result (\ref{spin-gg:LO}).

Cho and Leibovich \cite{cho-leibovich} have used the 
covariant projection method to calculate the cross sections for
$\psi$ and $\psi'$  in $p \bar p$ collisions, including not only the 
order-$\alpha_s^2$ parton processes but also the order-$\alpha_s^3$ 
processes of the form $i j \to c \bar c + k$. 
Our results (\ref{spin-qqbar}) and (\ref{spin-gg:LO})
for the cross sections from order-$\alpha_s^2$ parton processes
agree with theirs.  Cho and Leibovich 
determined the short-distance coefficients by matching
cross sections for producing color-octet $c \bar c$ pairs 
with vanishing relative momentum and in specific angular momentum 
states. 
This method gives the correct cross sections for unpolarized production,
but it fails in general for polarized production. 
In particular, as we show below, it fails to give the correct
short-distance coefficients for spin-triplet P-wave matrix elements.
In the matrix element (\ref{SS:83P}), the indices $i$ and $m$ can be
decomposed into contributions from total angular momentum 
$J = 0,1,2$ by using the identity
\begin{equation}
\delta_{i i'} \delta_{m m'}
\;=\; U^\dagger_{i \alpha} U^\dagger_{m \beta} \left( \sum_{J h} 
	\langle 1 \alpha ; 1 \beta | J h \rangle 
	\langle J h | 1 \alpha' ; 1 \beta' \rangle \right)
	U_{\alpha' i'} U_{\beta' m'},
\end{equation}
where a sum over $\alpha$, $\beta$, $\alpha'$, and $\beta'$ is implied.
A similar identity can be used to decompose the indices $j$ and $n$ 
into contributions from total angular momentum $J' = 0,1,2$.
The  matrix element (\ref{SS:83P}) is thereby expressed as a 
linear combination of matrix elements of the form
$\langle {\cal J}^\dagger(J,h) 
	{\cal P}_{\psi(\lambda)} {\cal J}(J',h') \rangle$,
where the operator ${\cal J}(J,h)$ creates a pointlike $c \bar c$ pair
with total angular momentum quantum numbers $J$ and $h$.
The projection operator 
${\cal P}_{\psi(\lambda)}$ projects onto states that contain a $\psi$ 
with helicity $\lambda$ plus soft hadrons.  If the soft hadrons have
total angular momentum  $J_S$, then the 
angular-momentum structure of the matrix element is
\begin{equation}
\sum_{\lambda_S} \langle J h | 1 \lambda; J_S \lambda_S \rangle
	\langle 1 \lambda; J_S \lambda_S | J' h' \rangle 
\label{CGcoeff}
\end{equation}
Rotational symmetry requires that $h = h'$ and that both $\underline J$ 
and $\underline J'$ lie 
in the irreducible decomposition of $\underline{1} \otimes \underline{J_S}$.
It does not require that $J = J'$, so there can be interference terms 
involving matrix elements with $J \neq J'$.  
If we sum (\ref{CGcoeff}) over the helicities
$\lambda$, the orthogonality relations for Clebsch-Gordan coefficients
imply that $J = J'$.  Thus the interference terms cancel upon summing over 
helicities.  The method used by Cho and Leibovich therefore
gives the correct cross sections for unpolarized production.
In the specific case of $g g \to c \bar{c}$, Cho and Leibovich
found that the only amplitudes for the production of 
spin-triplet P-wave  $c \bar c$ pairs
that are nonzero are $g g \to c \bar c(\underline{8},{}^3P_0)$ and 
$g g \to c \bar c(\underline{8},{}^3P_2, |h| = 2)$.
Since there is at most one nonzero amplitude for any given value of $h$,
there cannot be any interference terms. 
Thus their method gives the correct cross sections for polarized production
from the order-$\alpha_s^2$ parton processes. 
However, for the order-$\alpha_s^3$ parton processes
$i j  \to c \bar{c} + k$, there are nonvanishing amplitudes 
with  different values of $J$ and the same helicity $h$,
so the interference terms are nonzero. 
These interference terms must be taken into account in calculating the term
proportional to $\langle O_8^{\psi}(^3 P_0) \rangle$ in the inclusive cross
section for polarized $\psi$.  The only calculation thus far in which  
these interference terms have been correctly included is a calculation of the 
gluon fragmentation function for the production of a longitudinally polarized
$\psi$  by Beneke and Rothstein \cite{beneke-rothstein:1}.

Fleming and Maksymyk\cite{fleming-maksymyk} recently presented a method for
calculating cross sections for the production of unpolarized heavy quarkonium.
To determine the short-distance coefficients of the NRQCD matrix elements,
they matched cross sections for producing $c \bar c$ pairs 
with specific color and angular momentum quantum numbers.
The cross sections were expressed as integrals over the relative 3-momentum 
$\bf q$ of the $c \bar c$ pair, and they
used the nonrelativistic expansions of the Dirac spinors given in 
appendix A to expand the integrands as Taylor series in $\bf q$.
In general, their method would fail if applied to cross sections for 
polarized production, because it does not give the interference terms  
between parton processes that produce 
$c \bar c$ pairs with different total angular momentum $J$.
Fleming and Maksymyk applied their method to the production of unpolarized
$\psi$ from the parton processes $q\bar{q} \to c \bar{c}$ and 
$g g \to c \bar{c}$.  Our cross sections agree with theirs after 
summing over helicities.


\section{Summary}

The NRQCD factorization formalism \cite{BBL} is a powerful tool 
for analyzing the  production of
heavy quarkonium. Cross sections are factored  into 
short-distance coefficients which can be calculated using perturbative
QCD and long-distance
matrix elements which scale in a definite way with $v$. 
In this paper, we presented a general matching prescription for calculating
the short-distance coefficients in the inclusive cross sections 
for quarkonium states. 
Using this matching prescription, the cross sections are obtained in a form
that applies equally well to any quarkonium state.  The specific state
enters only in the reduction of the NRQCD matrix elements using rotational 
symmetry and heavy-quark spin symmetry.
For simplicity, we discussed the reduction of the NRQCD matrix elements 
only for the case of the $\psi$.
The generalization to other states will be presented elsewhere \cite{bc:next}.

Our approach has interesting implications for cross
sections for producing quarkonium states with definite polarization.
We showed that cross sections for polarized production can involve 
new matrix elements that do not contribute to cross sections for 
unpolarized production.  We also showed that there 
are interference terms involving the  production of $c\bar{c}$ states
with different total angular momentum $J$.
These interference terms must be taken into account in the
cross sections for polarized production from order-$\alpha_s^3$ 
parton processes.

Our method has a significant advantage over the covariant projection
method in that it can be straightforwardly generalized to $N$ space dimensions.
This is important, because it allows the use of dimensional regularization  
in calculations of radiative corrections.
There are potential inconsistencies in combining the covariant 
projection method 
with dimensional regularization for calculations involving orbital angular 
momentum $L =1$ or higher, because
the projections onto states with definite total angular momentum
$J$ are specific to 3 dimensions. 
Our method for calculating the short-distance coefficients of NRQCD
matrix elements involves matching Taylor expansions in the relative momentum, 
which can be readily calculated in $N$ dimensions.
After using renormalization to remove the poles in $1/(N-3)$ 
from the short-distance coefficients,  we can
specialize to $N = 3$ dimensions, and use 
rotational symmetry and heavy-quark 
spin symmetry to simplify the matrix elements.
Our approach thus allows the convenience of dimensional regularization 
to combined with the full power of the NRQCD factorization approach.
 
While this paper was being completed, 
Beneke and Rothstein \cite{beneke-rothstein} presented a paper 
that also points out that there are interference terms involving 
$c \bar c$ pairs with different total angular momentum.
They presented a thorough phenomenological analysis of 
the production in fixed-target hadron-hadron collisions of
$\psi$, $\psi'$, and 
$\chi_{cJ}$ via order-$\alpha_s^2$ parton processes.


\acknowledgements

This work was supported in part by the U.S.
Department of Energy, Division of High Energy Physics, under 
Grant DE-FG02-91-ER40690 and under Grant DE-FG02-91-ER40684.
We thank I. Maksymyk for pointing out an error in the preprint version 
of this paper.
\vfill\eject


\appendix

\section{Nonrelativistic expansion of spinors}

In this Appendix, we give the nonrelativistic expansions for 
the spinors of a heavy quark $c$ and antiquark $\bar c$ with arbitrary total
4-momentum $P$.  We assume that the relative 3-momentum ${\bf q}$
of the $c$ in the center-of-momentum (CM) frame  of the $c \bar c$ pair
is small compared to the quark mass $m_c$.  The 4-momenta
$p$ and $\bar p$ of the $c$ and $\bar c$ can be written
\begin{mathletters}
\begin{eqnarray}
p \;=\; \mbox{$1 \over 2$} P \;+\; L {\bf q} \;,
\\
\bar p \;=\; \mbox{$1 \over 2$} P \;-\; L {\bf q} \;,
\end{eqnarray}
\end{mathletters}
where $P$ is the total 4-momentum and $L$ is a Lorentz boost matrix.
>From the mass-shell conditions, $p^2 = \bar p^2 = m_c^2$, we have
$P \cdot L{\bf q} = 0$ and $P^2 = 4 E_q^2$, where 
$E_q = \sqrt{m_c^2 + {\bf q}^2}$. 
The components of the 4-momenta $P$ and $L {\bf q}$ in the CM frame 
of the pair are
\begin{mathletters}
\begin{eqnarray}
P^\mu \bigg|_{\rm CM} &\;=\;& ( 2 E_q , \; {\bf 0} ) \;,
\\
(L {\bf q})^\mu \bigg|_{\rm CM} &\;=\;& ( 0, \; {\bf q} ) \;.
\end{eqnarray}
\end{mathletters}
When boosted to an arbitrary frame in which
the pair has total 3-momentum ${\bf P}$, these 4-momenta are
\begin{mathletters}
\begin{eqnarray}
P^\mu &\;=\;&
\left( \sqrt{ 4 E_q^2 + {\bf P}^2} , \; {\bf P} \right) \;,
\\
(L {\bf q})^\mu &\;=\;& L^\mu_j \; q^j .
\end{eqnarray}
\end{mathletters}
The boost matrix $L^\mu_{\ j}$, which has one Lorentz index and one Cartesian
index, has components
\begin{mathletters}
\begin{eqnarray}
L^0_{\ j} &\;=\;& {1 \over 2 E_q} P^j \; , 
\\
L^i_{\ j} &\;=\;& \delta^{ij} - {P^i P^j \over {\bf P}^2}
	\;+\; {P^0 \over 2 E_q} {P^i P^j \over {\bf P}^2}  \;.
\end{eqnarray}
\end{mathletters}

The boost tensor $L^\mu_{\ i}$ has many useful properties.
Its contraction with the Lorentz vector $P$ vanishes: 
\begin{equation}
P_\mu L^\mu_{\ j} \;=\; 0 .
\label{id:PL}
\end{equation}
The contractions of two boost matrices in their Lorentz indices or in their
Cartesian indices have simple forms:
\begin{mathletters}
\begin{eqnarray}
g_{\mu \nu} L^\mu_{\ i} L^\nu_{\ j} &\;=\;& - \delta^{ij} ,
\label{id:gL}
\\
L^\mu_{\ i} L^\nu_{\ i} &\;=\;& - g^{\mu \nu} \;+\; {P^\mu P^\nu \over P^2} .
\label{id:LL}
\end{eqnarray}
\end{mathletters}
There are also simple identities involving contractions of boost matrices 
with the Levi-Civita tensors $\epsilon_{\mu \nu \lambda \rho}$
and $\epsilon^{i j k}$:
\begin{mathletters}
\begin{eqnarray}
\epsilon_{\mu \nu \lambda \rho} \; L^\mu_{\ i} L^\nu_{\ j} L^\lambda_{\ k}
&\;=\;& \epsilon^{ijk} \; {\hat P}_\rho ,
\\
\epsilon_{\mu \nu \lambda \rho} \; L^\mu_{\ i} L^\nu_{\ j} 
	{\hat P}^\lambda
&\;=\;& \epsilon^{ijk} \;  L_{\rho k},
\\
\epsilon_{\mu \nu \lambda \rho} \; L^\mu_{\ i} L^\nu_{\ j} 
&\;=\;& \epsilon^{ijk} \; 
	\left( {\hat P}_\lambda L_{\rho k} 
	- {\hat P}_\rho L_{\lambda k} \right),
\\
\epsilon_{\mu \nu \lambda \rho} \; L^\mu_{\ i} {\hat P}^\nu
&\;=\;& \epsilon^{ijk} \; L_{\lambda j}  L_{\rho k},
\label{epsid}
\\
\epsilon_{\mu \nu \lambda \rho} \; L^\mu_{\ i}
&\;=\;& \epsilon^{ijk} \; 
	\left( {\hat P}_\nu L_{\lambda j} L_{\rho k}
	+ {\hat P}_\lambda L_{\rho j} L_{\nu k} 
	+ {\hat P}_\rho L_{\nu j} L_{\lambda k} \right),
\\
\epsilon_{\mu \nu \lambda \rho} {\hat P}^\mu
&\;=\;& \epsilon^{i j k} \; L_{\nu i} L_{\lambda j} L_{\rho k},
\end{eqnarray}
\end{mathletters}
where ${\hat P}^\mu = P^\mu/\sqrt{P^2}$.
Our sign convention is $\epsilon^{0ijk} = - \epsilon_{0ijk} = \epsilon^{ijk}$.

The representation for gamma matrices that is most convenient 
for carrying out the nonrelativistic expansion of a spinor 
is the Dirac representation:
\begin{equation}
\gamma^0  \;=\; 
\left( \begin{array}{cc} 
	1 &  0  \\ 
	0 & -1 
	\end{array} \right) , 
\qquad
\gamma^i \;=\;  
\left( \begin{array}{cc} 
	     0    & \sigma^i \\ 
	-\sigma^i &    0        
	\end{array} \right) .
\end{equation}
In the CM frame of the $c \bar c$ pair, the spinors for the $c$ and 
the $\bar c$ are
\begin{mathletters}
\label{spin-rest}
\begin{eqnarray}
u(p) \Bigg|_{\rm CM} 
&\;=\;& {1 \over \sqrt{E_q + m_c}}
\left( \begin{array}{c} 
	(E_q + m_c) \; \xi \\ 
	{\bf q} \cdot \mbox{\boldmath $\sigma$} \; \xi 
	\end{array} \right) ,
\label{u-rest}
\\
v(\bar p) \Bigg|_{\rm CM}
&\;=\;& {1 \over \sqrt{E_q + m_c}}
\left( \begin{array}{c} 
	- {\bf q} \cdot \mbox{\boldmath $\sigma$} \; \eta \\
	(E_q + m_c) \; \eta
	\end{array} \right) .
\label{v-rest}
\end{eqnarray}
\end{mathletters}
Color and spin quantum numbers on the Dirac spinors
and on the 2-component Pauli spinors $\xi$ and $\eta$ are suppressed.
When boosted to a frame in which the pair has total 3-momentum ${\bf P}$,
the spinors for the $c$ and $\bar c$ are
\begin{mathletters}
\begin{eqnarray}
u(p) 
&\;=\;& {1 \over \sqrt{4 E_q (E_P + 2 E_q) (E_q + m_c)}}
\left( 2 E_q + \rlap/{\! P} \gamma_0 \right)
\left( \begin{array}{c} 
	(E_q + m_c) \; \xi \\ 
	{\bf q} \cdot \mbox{\boldmath $\sigma$} \; \xi 
	\end{array} \right) ,
\label{u-spinor}
\\
v(\bar p) 
&\;=\;& {1 \over \sqrt{4 E_q (E_P + 2 E_q) (E_q + m_c)}}
\left( 2 E_q + \rlap/{\! P} \gamma_0 \right)
\left( \begin{array}{c} 
	- {\bf q} \cdot \mbox{\boldmath $\sigma$} \; \eta \\
	(E_q + m_c) \; \eta
	\end{array} \right) .
\label{v-spinor}
\end{eqnarray}
\end{mathletters}
These spinors are normalized so that $\bar u u = - \bar v v = 2 m_c$ if the 
Pauli spinors are normalized so that $\xi^\dagger \xi = \eta^\dagger \eta = 1$.

There are 16 independent quantities that can be formed by sandwiching
Dirac matrices between $\bar u(p)$ and $v(\bar p)$.  They are
\begin{mathletters}
\begin{eqnarray}
\bar u(p) v(\bar p) &\;=\;& 
- 2 \; \xi^\dagger ({\bf q} \cdot \mbox{\boldmath $\sigma$}) \eta ,
\\
\bar u(p) \gamma^\mu v(\bar p) &\;=\;& L^\mu_{\ j} \;
\left( 2 E_q \; \xi^\dagger \sigma^j \eta \;-\; {2 \over E_q + m_c} \; q^j \;
	\xi^\dagger ( {\bf q} \cdot \mbox{\boldmath $\sigma$} ) \eta \right) \;,
\\
\bar u(p) \Sigma^{\mu \nu} v(\bar p) &\;=\;& 
( P^\mu L^\nu_{\ j} - P^\nu L^\mu_{\ j} ) 
\left( {i m_c \over E_q} \; \xi^\dagger \sigma^j \eta 
	\;+\; {i \over E_q(E_q + m_c)} q^j \; 
	\xi^\dagger ({\bf q} \cdot \mbox{\boldmath $\sigma$}) \eta \right)
\nonumber \\
&& \hspace{.5in}
\;-\; 2 \; L^\mu_{\ j} \; L^\nu_{\ k} \; \epsilon^{jkl} \; 
	q^l \; \xi^\dagger \eta,
\\
\bar u(p) \gamma^\mu \gamma_5 v(\bar p) &\;=\;& 
{m_c \over E_q} \; P^\mu \; \xi^\dagger \eta 
\;-\; 2 i \; L^\mu_{\ j} \;
	\xi^\dagger ({\bf q} \times \mbox{\boldmath $\sigma$})^j \eta ,
\\
\bar u(p) \gamma_5 v(\bar p) &\;=\;& 2 E_q \; \xi^\dagger \eta .
\end{eqnarray}
\end{mathletters}
>From these expressions, it is trivial to carry out the nonrelativistic
expansions in powers of ${\bf q}$.  For example, 
through linear order in ${\bf q}$, we have
\begin{mathletters}
\begin{eqnarray}
\bar u(p) v(\bar p) &\;=\;& 
- 2 \; \xi^\dagger ({\bf q} \cdot \mbox{\boldmath $\sigma$}) \eta ,
\\
\bar u(p) \gamma^\mu v(\bar p) &\approx& 
2 m_c \; L^\mu_{\ j} \; \xi^\dagger \sigma^j \eta \;,
\\
\bar u(p) \Sigma^{\mu \nu} v(\bar p) &\approx& 
i  \; ( P^\mu L^\nu_{\ j} - P^\nu L^\mu_{\ j} ) \; \xi^\dagger \sigma^j \eta 
\;-\; 2 \; L^\mu_{\ j} \; L^\nu_{\ k} \; \epsilon^{jkl} \; 
	q^l \; \xi^\dagger \eta,
\\
\bar u(p) \gamma^\mu \gamma_5 v(\bar p) &\approx& 
P^\mu \; \xi^\dagger \eta 
\;-\; 2 i \; L^\mu_{\ j} \;
	\xi^\dagger ({\bf q} \times \mbox{\boldmath $\sigma$})^j \eta ,
\\
\bar u(p) \gamma_5 v(\bar p) &\approx& 2 m_c \; \xi^\dagger \eta .
\end{eqnarray}
\end{mathletters}

\section{Production matrix elements}

In this Appendix, we define some of the standard NRQCD production matrix 
elements that were introduced in Ref. \cite{BBL}.  In order to establish 
the notation for the fields, we give 
the lagrangian for minimal NRQCD:
\begin{equation}
{\cal L}
\;=\; \psi^\dagger \, \left( iD_t + \frac{{\bf D}^2}{2M} \right)\, \psi
\;+\; \chi^\dagger \, \left( iD_t - \frac{{\bf D}^2}{2M} \right)\, \chi 
\;+\; {\cal L}_{QCD},
\label{Lheavy}
\end{equation}
where $\psi$ is the Pauli spinor field that annihilates a heavy quark,
$\chi$ is the Pauli spinor field that creates a heavy antiquark, and
${\cal L}_{QCD}$ is the usual QCD lagrangian for the gluons and the light 
quarks and antiquarks.

The production operators in Ref. \cite{BBL} were defined using a 
projection operator $a_H^\dagger a_H \equiv P_{M(\lambda)}^{\rm NR}$
defined by 
\begin{equation}
{\cal P}_{H(\lambda)}^{\rm NR} 
\;=\; \sum_S | H({\bf P} = 0,\lambda) + S \rangle 
	\langle H({\bf P} = 0,\lambda) + S | ,	
\label{P_HNR}
\end{equation}
where the states in the sum have the standard nonrelativistic normalization.
For example, the normalization of the quarkonium state is 
\begin{equation}
\langle H({\bf P}',\lambda') | H({\bf P},\lambda) \rangle 
\;=\;  (2 \pi)^3  \delta^3({\bf P} - {\bf P}') 
	\; \delta_{\lambda \lambda'}.	
\label{norm_HNR}
\end{equation}
Thus the projection operator (\ref{P_HNR}) has energy dimension -3.

The production operators of dimension 6 are
\begin{mathletters}
\begin{eqnarray}
{\cal O}^H_1({}^1S_0)
&\;=\;& \chi^\dagger \psi \; {\cal P}_{H(\lambda)}^{\rm NR} \; \psi^\dagger \chi,
\\
{\cal O}^H_1({}^3S_1) 
&\;=\;& \chi^\dagger \sigma^i \psi \; {\cal P}_{H(\lambda)}^{\rm NR} \;
	\psi^\dagger \sigma^i \chi,
\\
{\cal O}^H_8({}^1S_0) 
&\;=\;& \chi^\dagger T^a \psi \; {\cal P}_{H(\lambda)}^{\rm NR} \;
	\psi^\dagger T^a \chi,
\\
{\cal O}^H_8({}^3S_1)
&\;=\;& \chi^\dagger \sigma^i T^a \psi \; {\cal P}_{H(\lambda)}^{\rm NR} \;
	\psi^\dagger \sigma^i T^a \chi.
\end{eqnarray}
\end{mathletters}
Some of the production operators of dimension 8 are
\begin{mathletters}
\label{Odim8-prod}
\begin{eqnarray}
{\cal O}^H_1({}^1P_1)
&\;=\;& \chi^\dagger (-\mbox{$\frac{i}{2}$} \tensor{D}^i) \psi 
	\; {\cal P}_{H(\lambda)}^{\rm NR} \;
	\psi^\dagger (-\mbox{$\frac{i}{2}$} \tensor{D}^i) \chi ,
\\
{\cal O}^H_1({}^3P_{0}) 
&\;=\;& {1 \over 3} \; \chi^\dagger (-\mbox{$\frac{i}{2}$} \tensor{\bf D} 
		\cdot \mbox{\boldmath $\sigma$}) \psi
	\; {\cal P}_{H(\lambda)}^{\rm NR} \;
	\psi^\dagger (-\mbox{$\frac{i}{2}$} \tensor{\bf D} 
		\cdot \mbox{\boldmath $\sigma$}) \chi ,
\\
{\cal O}^H_8({}^1P_1)
&\;=\;& \chi^\dagger (-\mbox{$\frac{i}{2}$} \tensor{D}^i) T^a \psi 
	\; {\cal P}_{H(\lambda)}^{\rm NR} \;
	\psi^\dagger (-\mbox{$\frac{i}{2}$} \tensor{D}^i) T^a \chi ,
\\
{\cal O}^H_8({}^3P_{0}) 
&\;=\;& {1 \over 3} \; \chi^\dagger (-\mbox{$\frac{i}{2}$} \tensor{\bf D} 
		\cdot \mbox{\boldmath $\sigma$}) T^a \psi
	\; {\cal P}_{H(\lambda)}^{\rm NR} \;
	\psi^\dagger (-\mbox{$\frac{i}{2}$} \tensor{\bf D} 
		\cdot \mbox{\boldmath $\sigma$}) T^a \chi ,
\end{eqnarray}
\end{mathletters}
where $\psi^\dagger \tensor{\bf D} \chi 
	= \psi^\dagger {\bf D} \chi  - ({\bf D} \psi)^\dagger \chi$.

Note that the projection operator defined by (\ref{P_HNR}) differs
from the corresponding projection operator (\ref{P_H}), which is 
defined using states with the standard relativistic normalization.
At leading order in $v^2$, these operators differ simply by an
overall factor: 
${\cal P}_{H(\lambda)} \approx 4 m_c {\cal P}_{H(\lambda)}^{\rm NR}$.
Beyond leading order, the relation between the two projectors is 
complicated because 
the normalization differs for each term in the sum over soft states $S$.
For physical quantities, like the cross section in (\ref{match}),
the difference between the projection operators is compensated by 
the short-distance coefficients:
\begin{equation}
 \sum_{mn} C_{mn} \;
	\langle \chi^\dagger {{\cal K}'}_m^\dagger \psi \; 
	{\cal P}_{H(\lambda)} \;
	\psi^\dagger {\cal K}_n \chi \rangle 
\;=\; \sum_{mn} C_{mn}^{\rm NR} \;
	\langle \chi^\dagger {{\cal K}'}_m^\dagger \psi \; 
	{\cal P}_{H(\lambda)}^{\rm NR} \;
	\psi^\dagger {\cal K}_n \chi \rangle .	
\end{equation}

We now list the relations between scalar matrix elements defined with 
the projection operator (\ref{P_H}) and the matrix elements of the operators 
defined above.  At leading order in $\alpha_s$ and to leading order in $v^2$,
we have
\begin{mathletters}
\begin{eqnarray}
\langle \chi^\dagger \psi \; {\cal P}_{H(\lambda)} \; \psi^\dagger \chi \rangle
&\;=\;& 4 \; m_c \; {\cal O}^H_1({}^1S_0) ,
\\
\langle \chi^\dagger \sigma^i \psi \; {\cal P}_{H(\lambda)} \;
	\psi^\dagger \sigma^i \chi \rangle
&\;=\;& 4 \; m_c \; {\cal O}^H_1({}^3S_1) ,
\\
\langle \chi^\dagger (-\mbox{$\frac{i}{2}$} \tensor{D}^i) \psi \; 
	{\cal P}_{H(\lambda)} \;
	\psi^\dagger (-\mbox{$\frac{i}{2}$} \tensor{D}^i) \chi \rangle
&\;=\;& 4 \; m_c \; {\cal O}^H_1({}^1P_1), 
\\
\langle \chi^\dagger (-\mbox{$\frac{i}{2}$} \tensor{\bf D} \cdot 
		\mbox{\boldmath $\sigma$}) \psi \;
	{\cal P}_{H(\lambda)}^{\rm NR} \;
	\psi^\dagger (-\mbox{$\frac{i}{2}$} \tensor{\bf D} \cdot 
		\mbox{\boldmath $\sigma$}) \chi \rangle
&\;=\;& 12 \; m_c \; {\cal O}^H_1({}^3P_{0}) .
\end{eqnarray}
\end{mathletters}
The corresponding color-octet matrix elements are 
\begin{mathletters}
\begin{eqnarray}
\langle \chi^\dagger T^a \psi \; {\cal P}_{H(\lambda)} \; 
	\psi^\dagger T^a \chi \rangle
&\;=\;& 4 \; m_c \; {\cal O}^H_8({}^1S_0) ,
\\
\langle \chi^\dagger \sigma^i T^a \psi \; {\cal P}_{H(\lambda)} \;
	\psi^\dagger \sigma^i T^a \chi \rangle 
&\;=\;& 4 \; m_c \; {\cal O}^H_8({}^3S_1) ,
\\
\langle \chi^\dagger (-\mbox{$\frac{i}{2}$} \tensor{D}^i) T^a \psi \; 
	{\cal P}_{H(\lambda)} \;
	\psi^\dagger (-\mbox{$\frac{i}{2}$} \tensor{D}^i) T^a \chi \rangle
&\;=\;& 4 \; m_c \; {\cal O}^H_8({}^1P_1)  ,
\\
\langle \chi^\dagger (-\mbox{$\frac{i}{2}$} \tensor{\bf D} \cdot 
		\mbox{\boldmath $\sigma$}) T^a \psi \;
	{\cal P}_{H(\lambda)}^{\rm NR} \;
	\psi^\dagger (-\mbox{$\frac{i}{2}$} \tensor{\bf D} \cdot 
		\mbox{\boldmath $\sigma$}) T^a \chi \rangle
&\;=\;& 12 \; m_c \; {\cal O}^H_8({}^3P_{0}) .
\end{eqnarray}
\end{mathletters}
%


\end{document}